\documentclass[aps,showpacs,floatfix,a4paper,pre,preprint,groupedaddress]{revtex4}
\usepackage[dvips]{graphicx}
\usepackage{psfrag}
\usepackage{subfigure}
\usepackage{amsmath,color}
\usepackage{amsfonts}
\usepackage{mathrsfs} 
\usepackage{bm}
\newcommand{\ba}{\begin{eqnarray}}
\newcommand{\ea}{\end{eqnarray}}
\def\W{{\mathscr W}}
\def\R{{\mathscr R}}
\def\K{{\mathscr K}}
\def\F{{\mathscr F}}
\def\scov{\textit{\textbf D}_s}
\def\acov{\textit{\textbf W}_s}
\def\pt{{\partial}}
\def\Ds{\mathcal{D}_s}
\def\Dvs{\textit{\textbf D}_s}

\def\ev{{\bf e}}
\def\wv{{\bf w}}
\def\av{\textit{\textbf a}}
\def\nv{\textit{\textbf n}}

\def\bv{\textit{\textbf b}}

\def\kv{\textit{\textbf k}}
\def\kk{{\kappa}}

\def\wcon{w}
\def\wdis{\psi}
\def\tv{\textit{\textbf t}}
\def\vv{\textit{\textbf v}}

\def\gv{\textit{\textbf g}}
\def\gvc{\textit{\textbf g}^{(c)}}
\def\gvd{\textit{\textbf g}^{(d)}}

\def\Iv{\textit{\textbf I}}
\def\Gv{\textit{\textbf G}}

\def\Nv{\stackrel{\circ}{\nv}}
\def\Pv{\textit{\textbf P}}
\def\Iv{\textit{\textbf I}}

\def\Tv{{\bf T}}
\def\de{\delta}

\def\d{{\rm d}}

\def\tr{{\rm tr}}

\def\skw{{\rm skw}}
\def\grad{\nabla}
\def\grads{\nabla\hspace{-1mm}_s}

\def\curls{{\rm curl}_s}
\def\dvs{{\rm div}_s}
\def\esse{{\cal{S}}}

\def\bom{\vect{\omega}}
\def\bnu{\vect{\nu}}
\def\btau{\vect{\tau}}
\def\bsi{\vect{\sigma}}
\def\bsic{\vect{\sigma}^{(c)}}
\def\bsid{\vect{\sigma}^{(d)}}

\newcommand{\vect}[1]{\boldsymbol{\mathbf{#1}}}

\newcommand{\ot}{\otimes} 

\begin{document}
\title{Hydrodynamic theory for nematic shells: the interplay among curvature,  flow and   alignment}
\author{Gaetano Napoli}
\affiliation{Dipartimento di Ingegneria dell'Innovazione,
  Universit\`a del Salento, via per Monteroni, Edificio ``Corpo O'',
  73100 Lecce, Italy}
\author{Luigi Vergori} 
\affiliation{School of Mathematics and Statistics, University of Glasgow, University Gardens 15, G128QW Glasgow, UK }

\begin{abstract}
We derive the hydrodynamic equations for nematic liquid crystals lying on curved substrates. We invoke the Lagrange-Rayleigh variational principle to adapt the Ericksen-Leslie theory to  two-dimensional nematics in which a degenerate anchoring of the molecules on the substrate is enforced.  The only constitutive assumptions in this scheme concern the free-energy density, given by the two-dimensional Frank potential, and the density of dissipation which is required to satisfy appropriate invariance requirements. The  resulting equations of motion  couple the velocity field, the director alignment and the curvature of the shell. To illustrate our findings, we consider the effect of a simple shear flow on the alignment of a nematic lying on a cylindrical shell.    
\end{abstract}
\maketitle

A nematic liquid crystal is a fluid consisting of elongated  molecules that exhibit the tendency to align their major axes along a common direction. Such a preferred direction is usually described by a unit vector field $\nv$, called the \emph{nematic director}, which represents the local average orientation.
In the last two decades there has been an increasing interest in soft matter physics on  {\it nematic shells} \cite{Liang:2011, lopez:2011}. These shells, consisting  in  thin films of nematic liquid crystal deposited on  curved substrates,   are of fundamental interest because of their  suitability to study a rich variety of topological problems.  Furthermore, nematic shells  provide a promising route for generating colloids with controllable valence \cite{Nelson:2002, kralj:2011,  Liang:2012, Dhakal:2012}. The potential applications of nematic shells and their elegant formalism have produced a vivid research activity  \cite{vitelli:2006, bates:2008, kamien:2009,  naveprl:2012, Pairam:2013, Segatti:2014, Jesenek:2015, Koning:2015}. Most of the studies on nematic shells are addressed to understand the role of the shell geometry  on  the alignment  of nematics at equilibrium. However,
 while the static theory of nematic shells is gradually consolidating in the literature, a dynamic theory for two-dimensional curved nematics is still missing. 

From the dynamical point of view,  liquid crystals are complex non-Newtonian fluids whose continuum dynamical theory  is the result of the independent contributions by  Ericksen  and Leslie. We refer the reader  to \cite{Stewart:2004, Sonnet:2012} for an exhaustive and compendious treatise on the Ericksen-Leslie theory. The hydrostatic theory of nematics has been first obtained by
Ericksen  \cite{Ericksen:1962}  who reformulated the Frank theory  in the  general continuum mechanics framework. Ericksen \cite{Ericksen:1962} showed that the  stress tensor  is the sum of the usual hydrostatic isotropic term  and a {\it non-symmetric} contribution due to the anisotropy induced by the presence of a preferred direction.   Subsequently, broadening the constitutive assumptions on viscous dissipative actions,  Leslie \cite{Leslie:1968, Leslie:1968a}  derived a general dynamical theory that accounts for the fluid anisotropy and elastic stresses resulting from the spatial distortion of the director. In this model the dissipative dynamics is characterized by six viscosity coefficients.  
Few years later, Parodi \cite{Parodi:1970}  proved that  Leslie theory can be derived from a dissipation potential, and, exploiting the Onsager reciprocal relations, he was able to find a linear relation among the Leslie viscosity coefficients which reduces the number of independent coefficients to five. Very recently, basing on a novel hydrodynamic theory of nematic liquid crystals \cite{Biscari:2014}, Biscari {\it et al.} \cite{Biscari:2016} obtained a (nonlinear) equation  relating the Leslie viscosity coefficients.

A different  perspective is offered by Sonnet and Virga \cite{Sonnet:2001} who derived the Ericksen-Leslie equations for  dissipative fluids with a general microstructure by extending  the variational principle introduced by Rayleigh \cite{Rayleigh:1878} to describe dissipative discrete systems to continua. This derivation is valid for quite general constitutive models for the free-energy and dissipation densities provided that the latter satisfies appropriate invariance requirements. 
 In this paper we shall employ the  Lagrange-Rayleigh  variational principle  to obtain the hydrodynamic equations for a  2D nematics coating a fixed surface. Compared to the usual 3D nematics,  the number of degrees of freedom of a  nematic shell is reduced as  the center of mass of each molecule can move only on the substrate's surface (which involves two degrees of freedom) and the molecular axis can only  rotate rigidly around the normal to the substrate surface (so that its orientation can be described by  a single scalar parameter). Thus,  including the continuity equation in the system of governing equations, we expect to derive four scalar equations of motion instead of the six ones needed for a usual 3D nematics. 

Our variational approach is based on two constitutive ingredients: the free-energy density and the dissipation function. The free-energy  is given by the two-dimensional Frank-like potential that we derived in  \cite{nave:2012,naveprl:2012} and accounts for the interplay between the shell curvature  and the nematic alignment at equilibrium. On the other hand,  we here introduce  an appropriate model for the dissipation function which depends  on objective quantities. The equations of motion  we  obtain  clearly couple the nematic alignment, the {\it in-plane (covariant) strain rate tensor}  and the curvature of the shell.

According to the variational approach proposed in \cite{Sonnet:2012}, the equations of motion can be derived from the Lagrange-Rayleigh variational principle
\ba
\de \W = \de \R,
\label{variazionale}
\ea
 $\W$ being the total rate of work and $\R$  the Rayleigh dissipation functional. We refer the reader to the book by Sonnet and Virga \cite{Sonnet:2012} for a detailed discussion on the theoretical foundations and the pertinence of this principle. Hereinafter, we shall instead focus on the most appropriate  models for $\W$ and $\R$. 

We start by writing the total rate of work as the sum of three different contributions:
$
\W = \W^{(a)} - \dot \K - \dot \F,  
$
where $\W^{(a)}$ is the power of the external actions,  $\K$ is the kinetic energy, $\F$ is the free energy and the superimposed dot denotes differentiation with respect to time.

Next, we consider a  nematic  shell occupying  the surface $\esse$  with  unit normal $\bnu$, and denote $\pt \esse$ the boundary of $\esse$ and $\kv$ the unit outward normal to $\pt \esse$ in the tangent plane. The  power of the external actions results in the sum of  the power of the external forces acting on the material element of fluid and the power expended to make the molecules rotate:
\ba
  \W^{(a)}  = \int_{\mathcal{S}}\left({ \bv \cdot \vv  + \vect{\beta} \cdot \dot \nv}\right) \d a  + 
 \int_{\pt \esse}\left({\tv \cdot \vv  +   \vect{\gamma} \cdot \dot{\nv}}\right) \d l,
\label{wa}
\ea
where $\bv$ represents the surface density forces and $\tv$ the traction on the boundary, while $\vect{\beta}$ and $\vect{\gamma} $ are generalized body and contact force densities acting on $\nv$.

As the kinetic energy is concerned, beside the usual translational contribution, there is also a term  due to the rotation of the molecules. Nevertheless, because of the small moment of inertia of the molecules,  the rotational kinetic energy  can be ignored and thus  the  total kinetic energy reduces to
\ba
   \K = \frac{1}{2}\int_\esse \varrho \vv^2   \d a, 
 \label{kk}
 \ea
where $\varrho$ is the  mass density per unit of area.

Following \cite{naveprl:2012}, we assume that the free energy density $\wcon$ depends on the director field $\nv$ and its surface gradient $\grads \nv$ so that 
 \ba
  \F = \int_{\esse} \wcon(\nv, \grads \nv) \d a.
\label{ff}
\ea

The surface gradient is the differential operator $\grads\cdot=(\grad\cdot)\Pv$, where $\Pv = \Iv - \bnu \otimes \bnu$ denotes the projection operator onto the  plane tangent to $\esse$. The trace of surface gradient of a smooth vector  field $\wv$ gives the surface divergence, \emph{i.e.}, $\dvs\wv=\mathrm{tr}\grads\wv$, while twice the axial vector corresponding to the skew-symmetric part of $\grads\wv$ gives the surface curl of $\wv$: $\curls\wv=-\boldsymbol{\varepsilon}\grads\wv$, $\boldsymbol{\varepsilon}$ being the Ricci alternator. The surface gradient of $\wv$ is related to the covariant derivative $\Ds\wv$ through the identity $\Ds\wv=\Pv\grads\wv$ \cite{gurtin:1975}. Finally, the surface divergence of a second-order tensor field $\mathbf{\Tv}$ is the vector $\dvs \Tv$ with components $T_{ij,k}P_{jk}$.  

A 3D nematic liquid crystal 
dissipates energy through the fluid velocity $\vv$ and the local average angular velocity of the molecules. In the Ericksen-Leslie theory the dissipation functional
is a frame-indifferent quadratic form in  the strain rate tensor  (\emph{i.e.}, the symmetric part of the velocity gradient) and the {\it corotational} time derivative of $\nv$. 

\begin{figure}
	\centering
		\includegraphics[ width=8cm, keepaspectratio]{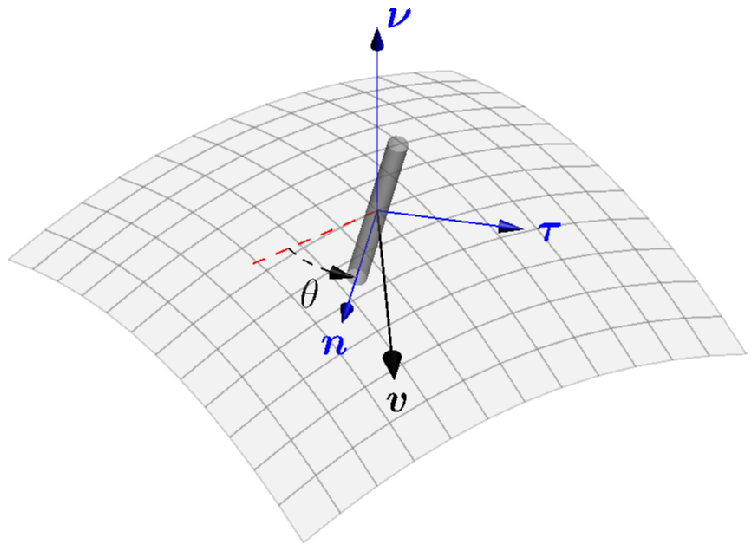}
		\caption{Darboux frame $\{\nv, \btau, \bnu\}$ on the nematic shell $\esse$. The vector $\vv$ is the velocity of the centre of mass of the molecule and $\theta$ is the angle contained by the director and the dashed line of curvature. \label{fig:darboux}}
\end{figure}

Unlike 3D  nematics, in nematic  shells the symmetric part of the velocity gradient and the corotational time derivative of the director  are not frame indifferent.  Instead, the appropriate frame indifferent tensorial quantity, linear in the velocity gradient, is the in-plane (covariant) strain tensor
$
 {\scov} = [\Ds \vv + (\Ds \vv )^T]/2,
 $
 and the appropriate frame-indifferent time derivative   of $\nv$ is the \emph{in-plane corotational time derivative}
\ba
\Nv \; =  \frac{\pt\nv}{\pt t}+(\Ds\nv)\vv - \acov \nv, 
\label{Nv}
\ea
where $\acov = [\Ds \vv - (\Ds \vv )^T]/2$ is the \emph{in-plane (covariant) vorticity tensor}.  The in-plane corotational time derivative $\Nv$ gives a measure of  the average angular velocity of the molecules contained in the material element relative to the regional angular velocity in which the material element is embedded. In fact, introduced the  {\it vorticity vector} $\boldsymbol{\varpi}=\curls\vv/2$ and denoting $\varpi_\nu$ its normal component,
$ 
\Nv\;=\left(\dot{\theta}-\bom\cdot\vv-\varpi_\nu\right)\bnu\times \nv,
$
where $\theta$ is the angle that the director forms with one of the principal directions (see Figure \ref{fig:darboux}) and $\bom$ is the vector which parametrizes the spin connection on $\esse$ \cite{nave:2012}.  The   local average angular velocity of the molecular axes  relative to the fluid is then $(\dot{\theta}-\bom\cdot\vv-\varpi_\nu)\bnu$. Consequently, the dissipation function can be taken of the form
\ba
\R = \int_{\esse} \wdis(\nv;\scov,\Nv) \d a,
\ea 
 $\wdis$ representing the density of dissipation. We may conclude  that, since both $\scov$ and $\Nv$ depend on the covariant derivative of the velocity gradient (and not on the surface gradient of $\vv$),  the extrinsic curvature of the shell does not affect directly the dissipative process.

The peculiarity of an in-plane liquid crystal flow with a degenerate anchoring stands in the kinematics.  Since $\esse$ is fixed, the admissible  velocity field $\vv$ is tangential. 
 In the sequel, we shall restrict our attention to  {\it inextensible flows}, {\it i.e.}, motions during which local area is preserved.  Within this assumption  the velocity is divergence-free, namely
\ba\label{inex}
\dvs  \vv = 0,
\ea
and, as a cosequence of this constraint and the equation of continuity, the mass density   is constant. Next, since $\vv$ is a tangential vector field satisfying the inextensibility constraint \eqref{inex}, the virtual velocity $\delta \vv$ is tangential and divergence-free. On the other hand, since the director is a unit vector lying in the tangent plane the infinitesimal variations $\de {\dot \nv}$ allowed by the constraints  point along the {\it conormal vector} $\btau= \bnu \times \nv$, that is only virtual rotations about the normal to $\esse$ are kinematically admissible.

In line with the Lagrangian spirit, the aforementioned kinematic restrictions  lead to a model with a reduced number of (scalar) equations compared to  the 3D Leslie-Ericken theory.  The variational principle \eqref{variazionale} yields two equations of motion:  the {\it fluid equation}
\ba
\varrho \av_s = -\grads \varsigma +  \Pv \dvs \bsi +  \bv_s,
\label{eq:fluid}
\ea
where $\av$ is the acceleration, $\bsi$ is the surface stress tensor,  $\varsigma$ is the Lagrange multiplier related to the inextensibility constraint \eqref{inex} and  the subscript $s$ appended to  vector fields denotes their projections onto the tangent plane; and the {\it director equation}
\ba
\vect{\tau} \cdot \dvs \Gv - g_{\tau}    + \beta_{\tau} = 0,
\label{eq:n}
\ea
where
\ba\label{GG}
\Gv=\left(\frac{\pt \wcon}{\pt \grads\nv}\right)\Pv, \quad \gvc = \frac{\pt \wcon}{\pt \nv}, \qquad  \gvd =  \frac{\pt \wdis}{\pt \Nv},
\ea
and  $g_{\tau}$ and $\beta_\tau$ are the components of $\gv = \gvc + \gvd$ and  $\vect{\beta}$ along the conormal vector $\vect{\tau}$.
The fluid equation \eqref{eq:fluid} gives rise to two scalar equations for the in-plane components of the velocity, while the director equation \eqref{eq:n} is the scalar equation for the angle $\theta$. The inextensibility constraint
\eqref{inex}
closes the system of governing equations as it allows the determination of the Lagrange multiplier $\varsigma$.

The stress tensor $\bsi$ is the sum of a constitutive elastic part 
\begin{subequations}\label{TensoreCauchy}
\ba
\bsic =  - (\grads \nv)^T \Gv,
\label{eri}
\ea
 and a viscous contribution
\ba
\bsid =  \frac{1}{2}\left(\nv \otimes \gvd_s - \gvd_s \otimes \nv \right) +  \Pv\frac{\pt \wdis}{\pt \scov}\Pv.
\label{les}
\ea
\end{subequations}
Notice that $\bsic$ is the two-dimensional analogue of the Ericksen stress tensor for usual 3D nematics  \cite{Ericksen:1962}. As in the 3D case, depending on the model for the free energy density $\wcon$, the conservative elastic part of the stress tensor $\bsic$ may fail to be symmetric.

The equations of motion are supplemented by  the boundary conditions
$
\tv_s = \bsi \kv$ and $\gamma_{\tau} = \vect{\tau} \cdot \Gv \kv$ on $\pt \esse$,
where $\tv_s$ is the tangential traction  and $\gamma_\tau$ is the component of the generalized boundary force along $\btau$.

Hereinafter, we consider specific models for the  free energy and dissipation densities. 

Frank \cite{frank:1958} was the first to derive the most general quadratic model  for the free energy density of nematic liquid crystals. Its adaptation to non-chiral nematics confined on curved surfaces is 
\ba
\wcon = \frac{k_1}{2}(\dvs \nv)^2 + \frac{k_2}{2}(\nv \cdot \curls \nv)^2 + \frac{k_3}{2} |\nv \times \curls \nv|^2
\label{franks}
\ea
where $k_1$, $k_2$ and $k_3$ are positive constants \cite{naveprl:2012}. Introducing the geodesic curvatures of the flux lines of $\nv$ and $\btau$,  denoted  $ \kk_{\nv}$ and $ \kk_{\btau}$, respectively, the normal curvature $c_\nv$ and the geodesic torsion $\tau_\nv$ \cite{docarmo}, 
equation (\ref{franks}) can be recast in the form
\ba \label{frankss}
\wcon= \frac{k_1}{2} \kk_{\btau}^2 + \frac{k_2}{2} \tau_\nv^2 + \frac{k_3}{2} \left(\kk_\nv^2 + c_\nv^2\right),
\ea
that is suited to an elegant and intuitive geometrical interpretation. Formula \eqref{frankss} highlights  the influence of  both the extrinsic and intrinsic curvatures of the substrate on the molecular alignment.  Specifically, the splay term tends to put the flux lines of $\btau$ along geodesics on $\esse$, while the term proportional to $\kk_\nv^2$ tries to align $\nv$ along geodesics. The twist energy favors instead the alignment of the flux lines of the director along  lines of curvature on $\esse$. Finally,  the term proportional to $c_\nv^2$ tends to align the flux lines of the director along the principal directions with minimum curvature (in modulus).  
 
From \eqref{GG} and \eqref{franks}
\begin{align}
\nonumber
\Gv &= k_1 \kappa_{\btau} \btau\ot\btau-k_2\tau_\nv\bnu\ot\btau\\
&+k_3(\kappa_\nv\btau\ot\nv+c_\nv\bnu\ot\nv)
\end{align}
and
\ba
g_\tau^{(c)} = (k_3-k_1)\kappa_\nv\kappa_{\btau}+(k_2-k_3)c_\nv\tau_\nv.
\ea
Within the {\it one constant approximation} ($k_1=k_2=k_3 \equiv k$), the free energy density \eqref{franks} reduces to $w = k |\grads \nv|^2/2$,
$\Gv = k \grads \nv$ and $g^{(c)}_\tau$ vanishes. 
 
The density,  derived here by analogy with the Ericken-Leslie 3D theory, must  obey the mirror symmetry 
$
\wdis(\nv;\scov, \Nv ) = \wdis(-\nv;  \scov, - \Nv).
$
Since neither $\scov$ nor $\Nv$ posses out-of-plane components, the most general model for $\wdis$,  quadratic in $\scov$ and $\Nv$, is
\ba
\wdis =  \frac{1}{2}\gamma_1 \Nv^2 + \gamma_2 \Nv \cdot \scov \nv + \frac{\gamma_3}{2} \tr \scov^2 + \frac{\gamma_4}{2} (\nv \cdot \scov \nv)^2,
\label{erre}
\ea
where $\gamma_i$ $(i=1,2,3,4)$ are viscosity coefficients.
A comparison with the three-dimensional analogue (\cite{Sonnet:2012}, page 176) shows that the dissipation function for nematic shells \eqref{erre} consists in four terms instead of five. Indeed, since $\scov$ is tangential and traceless the Hamilton-Cayley theorem  implies  that   $\nv\cdot \scov^2 \nv=2 \tr \scov$. 
Moreover, in virtue of the second law of thermodynamics, $\wdis$ is positive semi-definite, and so the viscosity coefficients satisfy the  inequalities 
\ba\label{gamma}
\gamma_1\geq0, \quad \gamma_3\geq0, \quad \gamma_2^2\leq2\gamma_1\gamma_3, \quad 2\gamma_3+\gamma_4\geq0.
\ea
We are now able to compute the dissipative quantities that are involved in the equations of motion.  From \eqref{TensoreCauchy}, \eqref{GG} and \eqref{erre} the viscous stress tensor is given by
\begin{align}
\bsid &=  \alpha_1 (\nv \cdot \Dvs \nv) \nv \otimes \nv + \alpha_2 \Nv \otimes \nv + \alpha_3 \nv \otimes \Nv \nonumber \\
&+ \alpha_4 \Dvs + \alpha_5 \skw(\nv \otimes \Dvs \nv),
\end{align}
where  
\begin{equation}\label{alpha's}
\left.\begin{array}{cc}
\alpha_1= \gamma_4, \quad 2 \alpha_2 = \gamma_2 - \gamma_1, \quad 2\alpha_3 = \gamma_1 + \gamma_2,\\
[2mm]
\alpha_4 = \gamma_3 \qquad \alpha_5 =  \gamma_2
\end{array}\right.
\end{equation}
are the Leslie viscosity coefficients, and
\begin{equation}\label{gvdfin}
\gvd = \gamma_1 \Nv + \gamma_2 \Dvs \nv.
\end{equation}
From \eqref{alpha's} we deduce  the Parodi-like identity $ \alpha_5 = \alpha_2 + \alpha_3$, while from \eqref{gamma} and \eqref{alpha's} we deduce that the Leslie viscosity coefficients satisfy the inequalities
\begin{equation}\label{alphadis}
\left.\begin{array}{cc}
\alpha_2\leq\alpha_3, \quad \alpha_4\geq0,  \quad \alpha_1+2\alpha_4\geq0,\\
[3mm]
 (\alpha_2+\alpha_3)^2\leq2\alpha_4(\alpha_3-\alpha_2).
\end{array}\right.
\end{equation}

To provide a basic understanding of the effect of the surface curvature on the director alignment, we consider a sample of nematic liquid crystal flowing on a cylindrical  shell  of radius $r$, height $h$ and parametrized through the local coordinates $\{\varphi,z\}$. We assume that the flow is laminar and the velocity of the form $\bar{\vv}(z)=  (\bar{v} z /h)\ev_\varphi $ ($\bar{v}>0$). It can be shown that $\bar{\vv}$ is a solution to the fluid equation whenever the classical viscosity $\alpha_4$ is much greater then the remaining Leslie viscosity coefficients.  
In the simplest possible scenario, the director depends only on time and so it can be  parametrized   as 
$
\nv =\cos\theta\ev_\varphi+\sin\theta\ev_z,
$
with $\theta=\theta(t)$. 

For the sake of illustration, we consider the free energy density within the one constant approximation. 
In this case the director  equation \eqref{eq:n} reads
\ba
\gamma_1 \theta_t  +  ( \alpha_3 \cos^2 \theta- \alpha_2 \sin^2 \theta)  \frac{\bar{v}}{h}-\frac{k}{2 r^2} \sin (2 \theta) =0.
\label{eq:alpha}
\ea
This equation shows that the director alignment is clearly affected by the extrinsic curvature of the substrate.

It is natural to proceed by looking for steady solutions ($\theta_t=0$) to \eqref{eq:alpha}. In doing so,  \eqref{eq:alpha} reduces to  the trigonometric equation
\ba
\frac \xi2 \sin (2 \theta)  + \alpha_2^* \sin^2 \theta -\alpha_3^*\cos^2\theta =0
\label{eq:a}
\ea
where $\xi=h^2/r^2$ and $\alpha_i^*=\alpha_i\bar{v}h/k$ ($i=2,3$).  For $\alpha_2\neq0$, equation  \eqref{eq:a} admits the solutions
\ba
\theta_\pm = \arctan\frac{-\xi \pm \sqrt{\xi^2 + 4\alpha_2^*\alpha_3^*}}{2\alpha_2^*},
\label{alphapm}
\ea
provided that $\alpha_2^*\alpha_3^*  \ge -\xi^2/4$. As expected, the curvature promotes the alignment of the director toward the cylinder generatrices (that are the directions of minimum curvature), while the flux tends to orient the molecules along the azimuthal direction. On the contrary, in the planar case $\xi$ vanishes and so \eqref{alphapm} admits a solution if and only if $\alpha_2^*\alpha_3^*$ is non-negative. Consequently, the {\it alignment angle}  depends only on the viscosity coefficients ratio $\alpha_3/\alpha_2$,
 $
 \theta_{\pm}^{(p)} = \pm \arctan \sqrt{\alpha_3/\alpha_2}.
 $ A  linear stability analysis reveals that in the cylindrical case $\theta_-$ is stable whereas $\theta_+$ is unstable. In the planar case instead the stability of the steady solutions $\theta_\pm^{(p)}$ depends on the signs of the viscosity coefficients $\alpha_2$ and $\alpha_3$. If $\alpha_2$ and $\alpha_3$ are both positive, $\theta_+^{(p)}$ is unstable and $\theta_-^{(p)}$ is stable. On the contrary, if  $\alpha_2$ and $\alpha_3$ are both negative, $\theta_+^{(p)}$ is stable and $\theta_-^{(p)}$ is unstable. If $\alpha_3=0$, then the only steady solution $\theta^{(p)}=0$ is neutrally stable. It is worth noting that the stability results in the planar and cylindrical cases  match each other because, as $\xi\rightarrow0$, $\theta_+\rightarrow\theta_+^{(p)}$  if $\alpha_2$ and $\alpha_3$ are both positive,  $\theta_+\rightarrow\theta_-^{(p)}$ if $\alpha_2$ and $\alpha_3$ are both negative, and $\theta_\pm\rightarrow 0$ if $\alpha_3=0$.

\begin{widetext}

\begin{figure}[h]
	\centering
	\psfrag{x}{{$|\alpha^*_2|$}}
	\psfrag{a1}{{$\alpha_2/\alpha_3$}}
	\psfrag{t}{{\hspace{-0.2cm}$\theta$ }}
	\psfrag{data1}{\tiny{$\xi=0$}}
	\psfrag{data2}{\tiny{$\xi=0.5$}}
	\psfrag{data3}{\tiny{$\xi=1$}}
	\psfrag{data4}{\tiny{$\xi=2$}}
	\psfrag{y1}{\tiny{$\xi/\alpha_3^*=0$}}
	\psfrag{y2}{\tiny{$\xi/\alpha_3^*=0.5$}}
	\psfrag{y3}{\tiny{$\xi/\alpha_3^*=1$}}
	\psfrag{y4}{\tiny{$\xi/\alpha_3^*=2$}}
	\psfrag{v1}{\tiny{$\xi/|\alpha_3^*|=0$}}
	\psfrag{v2}{\tiny{$\xi/|\alpha_3^*|=0.5$}}
	\psfrag{v3}{\tiny{$\xi/|\alpha_3^*|=1$}}
	\psfrag{v4}{\tiny{$\xi/|\alpha_3^*|=2$}}
	\subfigure[\, $\alpha_3> 0$.\label{fig:a3pos}]
		{\includegraphics[ width=5.7cm, keepaspectratio]{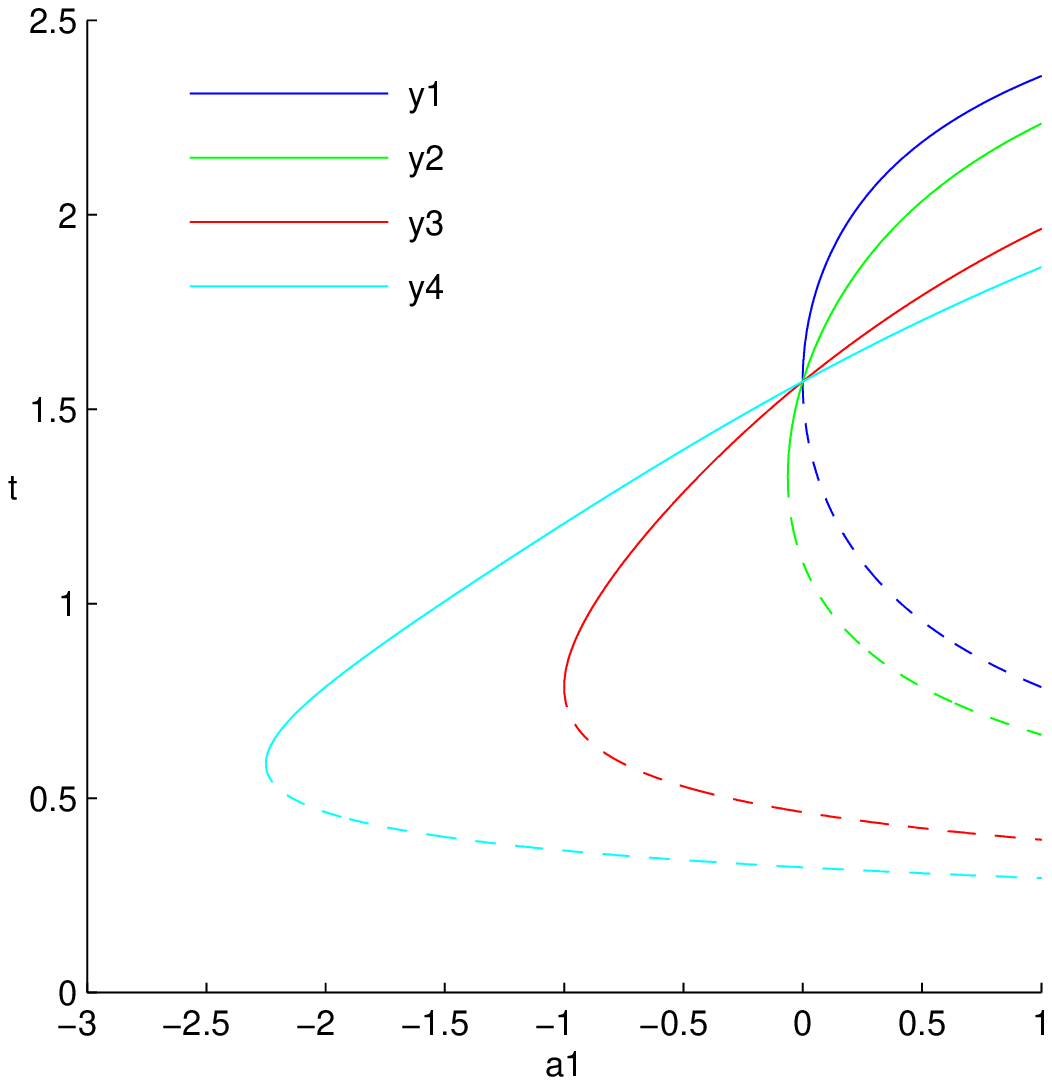}}
		\subfigure[\, $\alpha_3< 0$.\label{fig:a3neg}]
		{\includegraphics[ width=5.7cm, keepaspectratio]{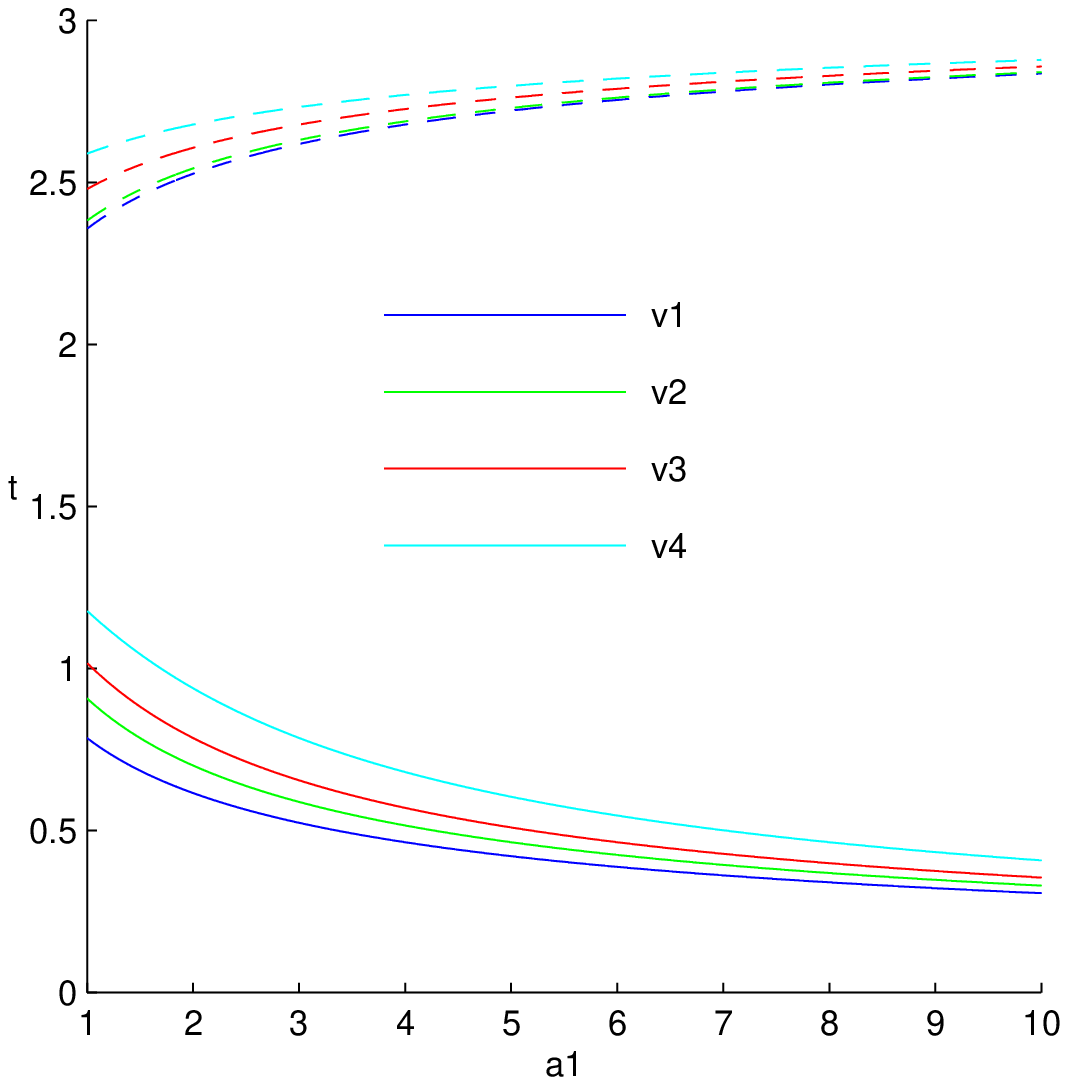}}
		\subfigure[\, $\alpha_3=0$.\label{fig:a30}]
		{\includegraphics[ width=5.7cm, keepaspectratio]{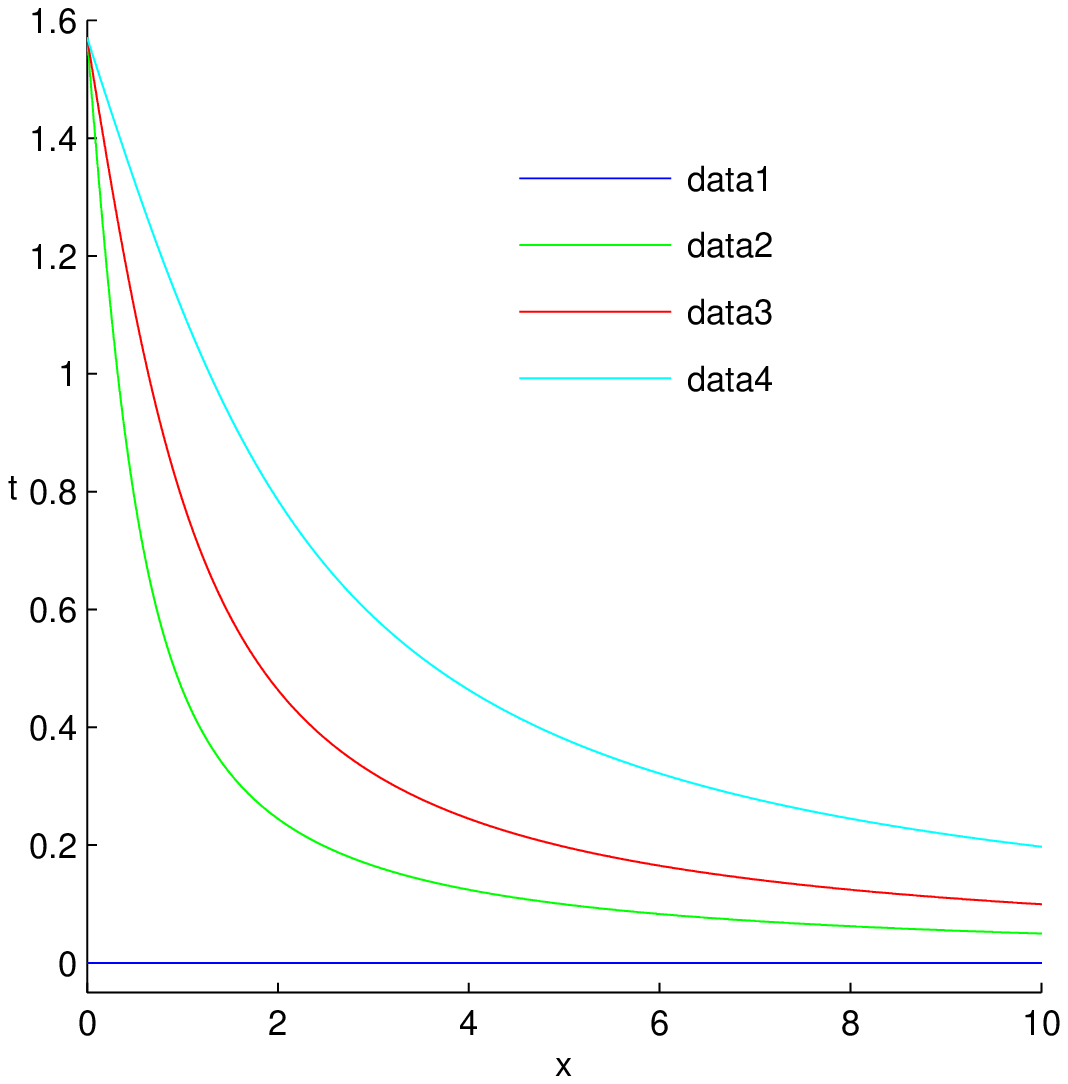}}
\caption{Alignment angles as functions of the Leslie viscosity parameters $\alpha_2$ and $\alpha_3$ for  cylindrical and planar geometries. The solid lines represent linearly stable steady solutions,  the dashed lines  unstable alignments. \label{steady}}
\end{figure}

 \end{widetext}

In the special case in which $\alpha_2$ vanishes, in the cylindrical case the  steady solutions to \eqref{eq:alpha} are
\begin{equation}
\theta_1=\frac{\pi}{2},\quad \theta_2=\arctan\frac{\alpha_3\bar{v}r^2}{kh},
\end{equation}
$\theta_1$ being linearly stable and $\theta_2$ linearly unstable. For the sake of completeness, in the planar case the only steady solution to \eqref{eq:alpha},  $\theta_1=\theta_2=\pi/2$, is neutrally stable. We finally observe that, as $\alpha_2\rightarrow0$, $\theta_-\rightarrow\theta_1$ and $\theta_+\rightarrow\theta_2$ (see Figure \ref{steady}). 

 In summary, we have obtained a Leslie-Ericksen-like theory for nematic shells. The geometrical constraint   allows not only the reduction of  the degrees of freedom with respect to the 3D theory, but reduces also  the number of viscosity coefficients.   The complete description of the dynamical problem needs only  four scalar equations: two related to the in-plane fluid motion, one for the determination of the Lagrange multiplier associated with  the inextensibility constraint, and one for the evolution of the director. However, the main result is the coupling among the nematic alignment, the fluid flow and the intrinsic and extrinsic curvatures of the shell. { We have shown that, unlike the free-energy (Frank-like potential) where both the intrinsic and the extrinsic curvatures are concerned, the density of dissipation (Leslie-Ericksen-like functional) involves only covariant quantities as a consequence of the invariance requirements.}
The interplay among flow, nematic texture and curvature is well highlighted in the example provided.
 
 The potential applications of this model are varied. For example, the motion of topological defects (interaction with the backflow), their enucleation or disappearance, as well as their stability, may be carried out by means of numerical simulations employing the equations derived here. In addition, our model is the first attempt towards more complex dynamics. { On one hand, our variational scheme can be easily extended to the case of active nematics by simply modifying the dissipation functional $\R$  to include the presence of motors. On the other hand, the model can be easily generalized to take into account  other kinematic descriptors such as, for instance, the Landau-de Gennes order-tensor.} 
 

\end{document}